\documentclass[journal=apchd5,manuscript=letter,layout=traditional]{achemso}

\usepackage[version=3]{mhchem}
\usepackage{amssymb}
\usepackage{gensymb} 
\usepackage{siunitx}
\usepackage{textcomp}
\usepackage{color,soul}

\author{Tianyu Fang}
\affiliation{Institute of Physics, University of Kassel, Heinrich-Plett-Stra{\ss}e 40, 34132 Kassel, Germany}
\author{Florian Elsen}
\affiliation{Institute of Physics, University of Kassel, Heinrich-Plett-Stra{\ss}e 40, 34132 Kassel, Germany}
\author{Nick Vogeley}
\affiliation{Institute of Physics, University of Kassel, Heinrich-Plett-Stra{\ss}e 40, 34132 Kassel, Germany}
\author{Daqing Wang}
\affiliation{Institute of Physics, University of Kassel, Heinrich-Plett-Stra{\ss}e 40, 34132 Kassel, Germany}
\email{daqing.wang@uni-kassel.de}

\title{Optical imaging and tracking of single molecules in ultrahigh vacuum}


\begin{document}



\begin{abstract}
 Molecule-surface interaction is key to many physical and chemical processes at interfaces. Here, we show that the dynamics of single molecules on a surface under ultrahigh vacuum can be resolved using fluorescence imaging. By adapting oil-immersion microscopy to a thin vacuum window, we measure the surface adsorption, translational and rotational diffusion of single perylene molecules on a fused silica surface with high spatial and temporal resolutions. {Time-dependent measurements of the fluorescence signal allow us to deduce two characteristic decay time scales, which can be explained through a simplified model involving two adsorption states and five energy levels}. The system presented in this work combines fluorescence imaging with essential ingredients for surface science and promises a platform for probing single molecule-surface interactions in highly defined conditions.
\end{abstract}

\section*{Introduction}
Physical and chemical processes at interfaces are fundamental to a wide range of applications covering surface catalysis\cite{Somorjai917}, optoelectronics\cite{Chen2021}, life science\cite{Almeida2021}, and quantum technologies\cite{Su2016,Gaita-Arino2019}. The ability to resolve individual molecules and track their dynamics on well-defined surfaces is crucial for the understanding of these processes. {Motivated by these interests, molecule-surface interaction has been studied in different scientific contexts and employing different technologies. Scanning probe techniques such as atomic force microscopy\cite{deOteyza2013,Riss2016} and scanning tunneling microscopy\cite{Wong2007,Civita2020} are proven tools for imaging and manipulating single molecules on surfaces with spatial resolutions down to the atomic scale. Harnessing the nanocavity plasmon modes supported by the tip-substrate geometry, scanning probe techniques can be further combined with laser excitation, allowing optical microscopy and spectroscopy through the detection of fluorescence\cite{Imada2021, Yang2020} or tip enhanced Raman scattering\cite{Prabhat2017, Imada2021}. Tip- and surface-enhanced Raman scattering provide rich spectroscopic information on the vibrational signatures of molecules and have become important tools for sensitive chemical and biological analysis\cite{Langer2020,Schultz2020}. Alternatively, interactions of small molecules with surfaces can also be probed through molecule beam scattering experiments in the gas phase\cite{Libuda2005}. In these measurements, scattered or desorbed molecules from surfaces are normally detected through photoionization and time-of-flight measurements. Information on the residing time, scattering velocity and activation energy for desorption can be obtained\cite{Harding2017}. Scanning probe and beam scattering experiments are typically operated under ultrahigh vacuum (UHV) and with well-defined surfaces. In contrast, single-molecule fluorescence imaging techniques have been widely applied to surfaces under ambient conditions, such as cell membranes\cite{Li2017}, and chemically relevant surfaces including porous materials\cite{Zurner2007,Maris2021} and catalyst structures\cite{Ruehle2012,Hendriks2017}. In combination with single-molecule localization methods\cite{Weisenburger2015}, fluorescence imaging allows wide-field visualization of molecular diffusions with simultaneously high temporal and spatial resolutions\cite{Lelek2021}.}

While optical microscopy of single molecules is widely applied in studying dynamic processes, a prerequisite is efficient photon collection\cite{Enderlein1997}. The fluorescence signal from a single molecule in a detection time window is ultimately limited by its excited-state lifetime. In applications under ambient conditions, efficient photon collection is commonly achieved through the use of oil-immersion objectives with high numerical apertures (NA). As a result, spatial localization precision of $\sim$10\,nm and time resolution of $\gtrsim$1000 frames per second are within reach\cite{Weisenburger2015,Lelek2021}. Implementing similar constructions in UHV settings is challenging due to the outgassing of the immersion oil and the imaging optics. Moreover, typical high-NA objectives have working distances in the order of 100\,{\micro}m, leading to increased complexities in positioning and vibration control inside the UHV chamber. These challenges can be circumvented by using a thin vacuum window, that allows a high-NA objective to image its interior surface. In the past decades, various types of thin windows have been reported for specialized applications. Imaging single molecules on the inner surface of a thin window was employed for spatially-resolved detection of matter-wave diffraction\,\cite{Juffmann2012,Brand2015}. Thin windows were also used for electron microscopy of samples in air and liquid\,\cite{Williamson2003,Han2016} and imaging cold atoms in an optical lattice\,\cite{Edge2015}. 

Here, we show that single-molecule dynamics on a surface under UHV can be studied with spatial and temporal resolutions at the same level with experiments under ambient conditions. This is enabled by the combination of oil-immersion microscopy with a home-made $150$\,{\micro}m-thin vacuum window.
\section*{Experimental setup}
\begin{figure}[h]
\centering\includegraphics[width=8.0cm]{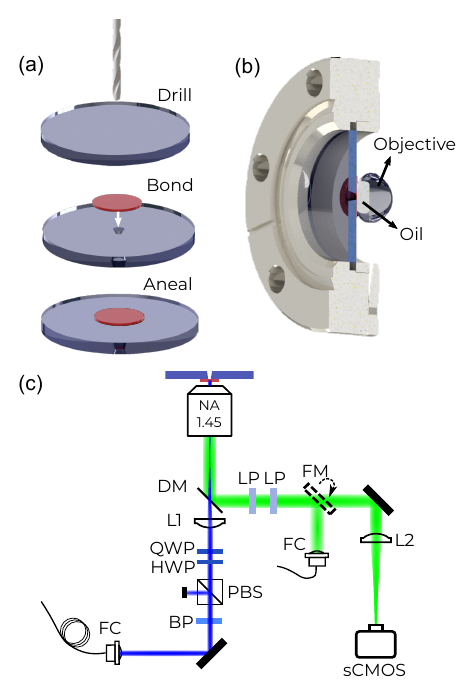}
\caption{(a) Fabrication of a thin vacuum window. Top: a 2 mm-diameter hole is drilled at the center of a fused silica window; Middle: a 150\,{\micro}m-thin coverslip is attached to the window through optical-contact bonding; Bottom: the assembly is annealed to form covalent bonds between the two pieces. (b) The window is mounted on a Conflat flange viewport with the coverslip facing the {atmospheric} side. An oil-immersion objective under ambient conditions images the inner surface of the coverslip. (c) Optical setup. FC: fiber coupler; BP: bandpass filter; PBS: polarization beamsplitter; HWP: halfwave plate; QWP: quarterwave plate; DM: dichroic mirror; LP: longpass filter; FM: flip mirror; L1, L2: $f=250$\,mm lenses; sCMOS: scientific complementary metal–oxide–semiconductor sensor. An objective with NA=1.45 is used for imaging.}
\label{fig_setup}
\end{figure}

To fabricate a vacuum window suited for oil-immersion microscopy, we start with a 4\,mm-thick fused-silica window plate, and drill a 2\,mm-diameter hole through its center using a diamond-coated drill head, as illustrated in Fig.\,\ref{fig_setup}(a). The window is then cleaned in acetone, methanol, piranha consecutively and rinsed with ultra-pure water. {A 150\,{\micro}m-thin fused silica coverslip, cleaned through the same procedure,} is brought into optical contact with the window through van der Waals bonding\,\cite{Mawatari2012book}, with its center matched to the hole on the window. The assembly is then transferred into a muffle furnace, annealed at 950\,$\degree$C, and under atmospheric pressure for about 8 hours. During this process, covalent bonds are developed between the two surfaces\,\cite{Ploessl1999} and the resulting piece becomes vacuum tight. The window is then installed on a demountable Conflat viewport (VPD-1.19, Accu-Glass Products Inc.), with the bonded coverslip facing the {atmospheric} side [see Fig.\,\ref{fig_setup}(b)]. The resulted viewport has a clear aperture of 30\,mm and the central bonded area has a vacuum safety factor\,\cite{Yoder2018book} of 10. 

The viewport is fixed onto a two-dimensional translation stage and connected to a vacuum chamber via a corrugated bellow. {The vacuum chamber is evacuated by a turbomolecular pump with a pump speed of 240\,L/s. The measurements reported in this work were performed in the pressure range of $7.0\times10^{-9}$\,mbar to $3.0\times10^{-8}$\,mbar. The pressure can be further reduced through a bakeout of the vacuum chamber with the viewport to 120\,\celsius, leading to $1.3\times10^{-9}$\,mbar.} The translation stage holding the vacuum window facilitates the transverse adjustment of the latter. An oil-immersion microscope objective (M Plan Apochromat 100x, Olympus; NA=1.45) is positioned to image the inner surface of the window. Low-autofluorescence immersion oil is applied between the objective and the coverslip, as displayed in Fig.\,\ref{fig_setup}(b). 

The light emitted by a diode laser at 405\,nm is used to illuminate the inner surface of the coverslip. The laser is first coupled out from a fiber coupler, spectrally cleaned by a bandpass filter {(HC Laser Clean-up MaxDiode 405/10, IDEX/Semrock)} and sent through a polarization beamsplitter and a pair of half- and quarter-waveplates. Before entering the objective, a $f=250$\,mm lens focuses the light to achieve a wide-field illumination, see Fig.\,\ref{fig_setup}(c). Perylene molecules ($\geq$99\%, Sigma Aldrich) are used for the experiment. The molecules are brought into the gas phase through laser ablation on a target of perylene microcrystals located 80\,cm upstream of the observation window. {In between measurements, the coverslip can be cleaned with strong laser illumination, without the need of disassembling the window}. The fluorescence signal of the molecules is collected by the same microscope objective, separated from the illumination light by a {dichroic} mirror, filtered by two longpass filters {(430/LP BrightLine and 470/100 BrightLine HC, IDEX/Semrock)} and imaged on a scientific Complementary Metal Oxide Semiconductor (sCMOS; Prime BSI, Teledyne Photometrics) camera.  Alternatively, the fluorescence light can be coupled to a multimode fiber and guided to a spectrometer.

\section*{Oil-immersion microscopy of single molecules}
\begin{figure}[t]
\centering\includegraphics[width=8.3cm]{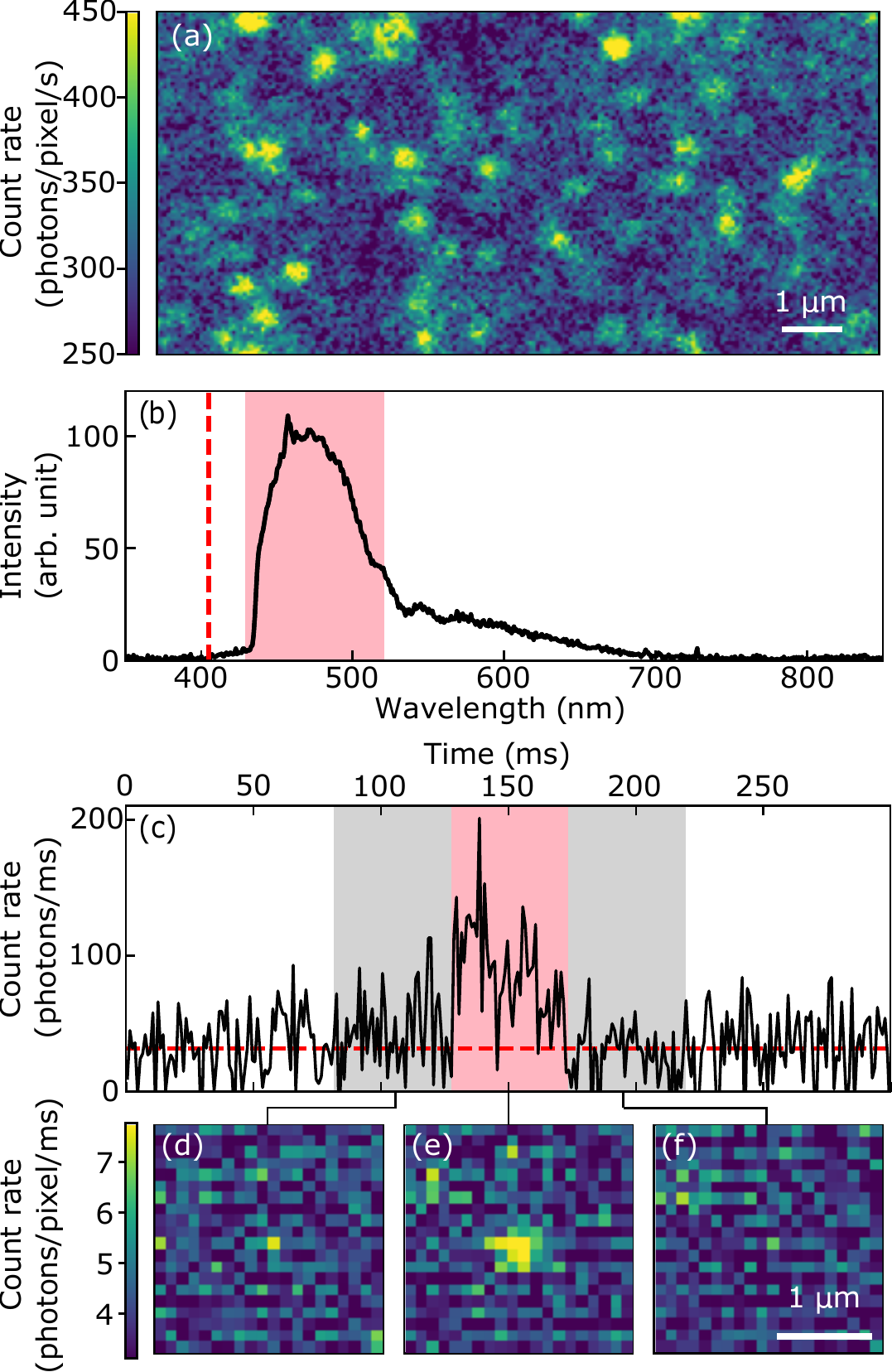}
\caption{(a) Wide-field fluorescence image of perylene molecules on the inner surface of the window. { (b) Emission spectrum of perylene molecules on fused silica surface. The red dashed line stands for the excitation wavelength and the red shaded area denotes the window of detection.} (c) The black curve shows the fluorescence intensity of an area of interest as a function of time. The dashed red line displays the average background level. The red shaded area indicates the adsorption event of a single molecule. (d-f) Averaged fluorescence images of the frames marked by {shaded} areas in (c).}
\label{fig_imaging}
\end{figure}

Figure\,\ref{fig_imaging}(a) displays a wide-field fluorescence image of the inner surface of the fused silica window. {Single perylene molecules appear as bright spots with a coverage of about 2 molecules per {\micro}m$^2$}. The image was acquired with an illumination intensity of $600$\,W/cm$^2$ and an exposure time of one second. {An emission spectrum of perylene molecules recorded with a spectrometer (QEPro, Ocean Insight) is displayed in Fig.\,\ref{fig_imaging}(b). The dashed red line indicates the excitation wavelength of 405\,nm. The red shaded area illustrates the detection window of 430\,nm to 520\,nm, defined by the interference filters. Compared to perylene spectra obtained in solutions \cite{Berlman2012handbook} and matrices\cite{Zondervan2003}, the multi-peak vibrational bands are not resolved. This can be a result of inhomogeneous broadening due to interactions with the surface.} The camera signal in Fig.\,\ref{fig_imaging}(a) was converted to photon counts per pixel by accounting for a digitization factor and a weighted quantum efficiency of 0.88 in the spectral range of detection.

The efficient photon collection with the oil-immersion objective allows the detection of single-molecule adsorption events with high temporal resolution. The black curve in Fig.\,\ref{fig_imaging}(c) displays the photon counts as a function of time from an area formed by $8\times8$ camera pixels. The dashed red line indicates the background level. The adsorption of a single molecule is visible in the time window of 130\,ms to 170\,ms, indicated by the red shaded area. The data were acquired at 979 frames per second with an exposure time of 1\,ms for each frame. Note that the black curve extends slightly to negative values due to the inhomogeneity of the pixel responses and the subtraction of a mean digitization offset. Figure\,\ref{fig_imaging}(d) displays the averaged camera image in a time window before the adsorption event, indicated by the grey shaded area in Fig.\,\ref{fig_imaging}(c). Image of the same area during the adsorption of a single molecule is shown in Fig.\,\ref{fig_imaging}(e). After 170\,ms, the signal returns to the background level, indicating that the molecule was either desorbed from the surface or suffered photobleaching [see also Fig.\,\ref{fig_imaging}(f)]. A $2\times2$ pixel binning was applied during the acquisition of these images.
\begin{figure}[ht]
\centering\includegraphics[width=6.5cm]{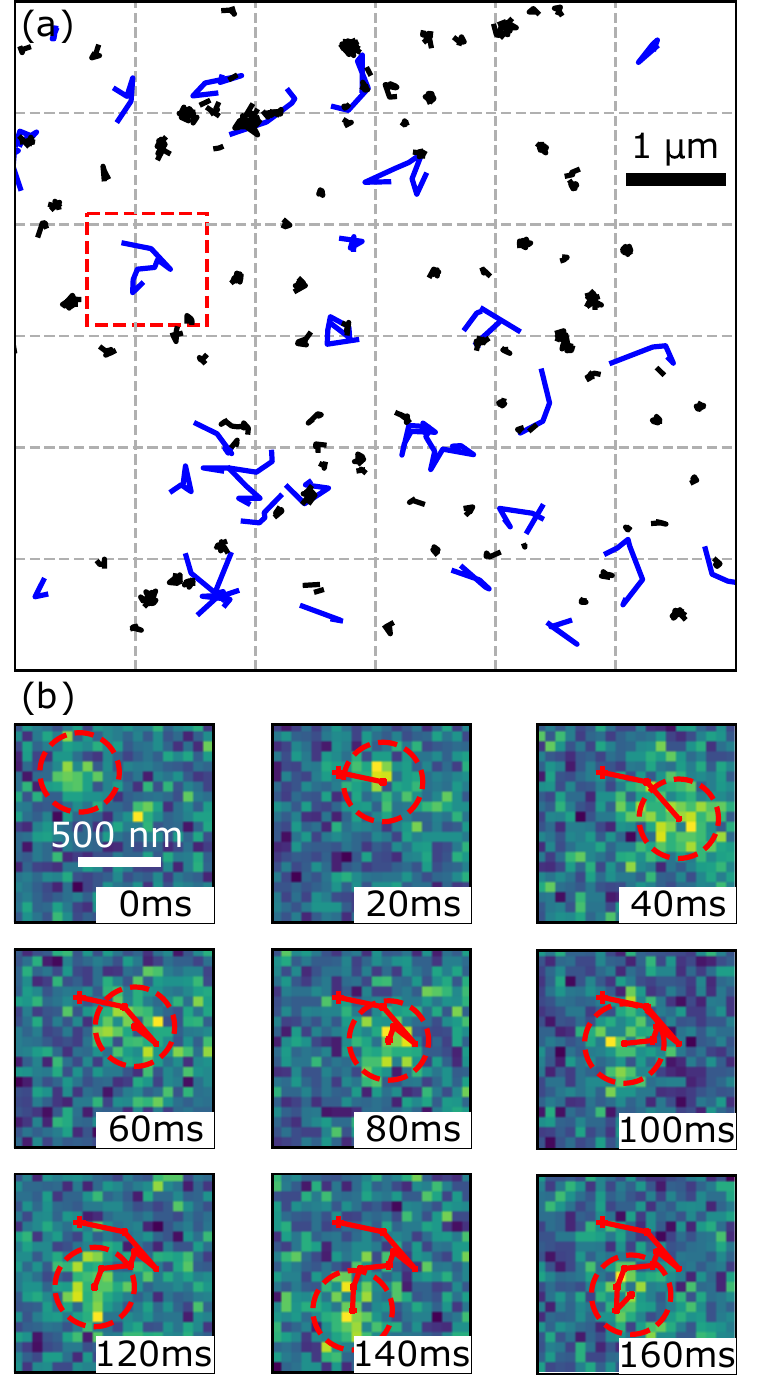}
\caption{(a) Trajectories of single perylene molecules identified during an observation time of two seconds. Black and blue trajectories indicate the molecules with low and high mobilities. (b) Consecutive camera images of the molecule marked by the red rectangular {box} in (a).}
\label{fig_diff}
\end{figure}

\section*{Translational and rotational diffusion}
Having the capability to image single adsorbed molecules, we turn to investigate their translational diffusion on the fused silica surface. To do so, we first identify individual molecules in each frame using the software TrackMate\cite{Tinevez2017}. For each identified molecule, a maximum likelihood estimation procedure is applied with the point-spread function (PSF) of the system as the model function. The resulting uncertainty of the center position yields the localization precision\cite{Weisenburger2015}. With 20\,ms exposure time on the camera, we achieve an averaged localization precision of $\pm$24\,nm. Localized single molecules in consecutive frames are linked based on their mutual distances\cite{Crocker1996}. Figure\,\ref{fig_diff}(a) displays the trajectories of 120 molecules identified during two seconds of observation time. We find that the majority of the molecules showing persistent fluorescence signal are strongly confined in position, as indicated by the black trajectories. These molecules are likely tightly bound to their local adsorption sites, and their coupling to the surface quenches the long-lived internal states efficiently, yieldings stable fluorescence signals. In contrast, we also identified molecules with significantly higher mobility, as indicated by the blue trajectories in Fig.\,\ref{fig_diff}(a). These molecules show reduced fluorescence stability and shorter residence times on the surface. Figure\,\ref{fig_diff}(b) displays consecutive camera frames of a molecule indicated by the dashed red square in Fig.\,\ref{fig_diff}(a), showing clear translational diffusion.

\begin{figure}[ht]
\centering\includegraphics[width=7.9cm]{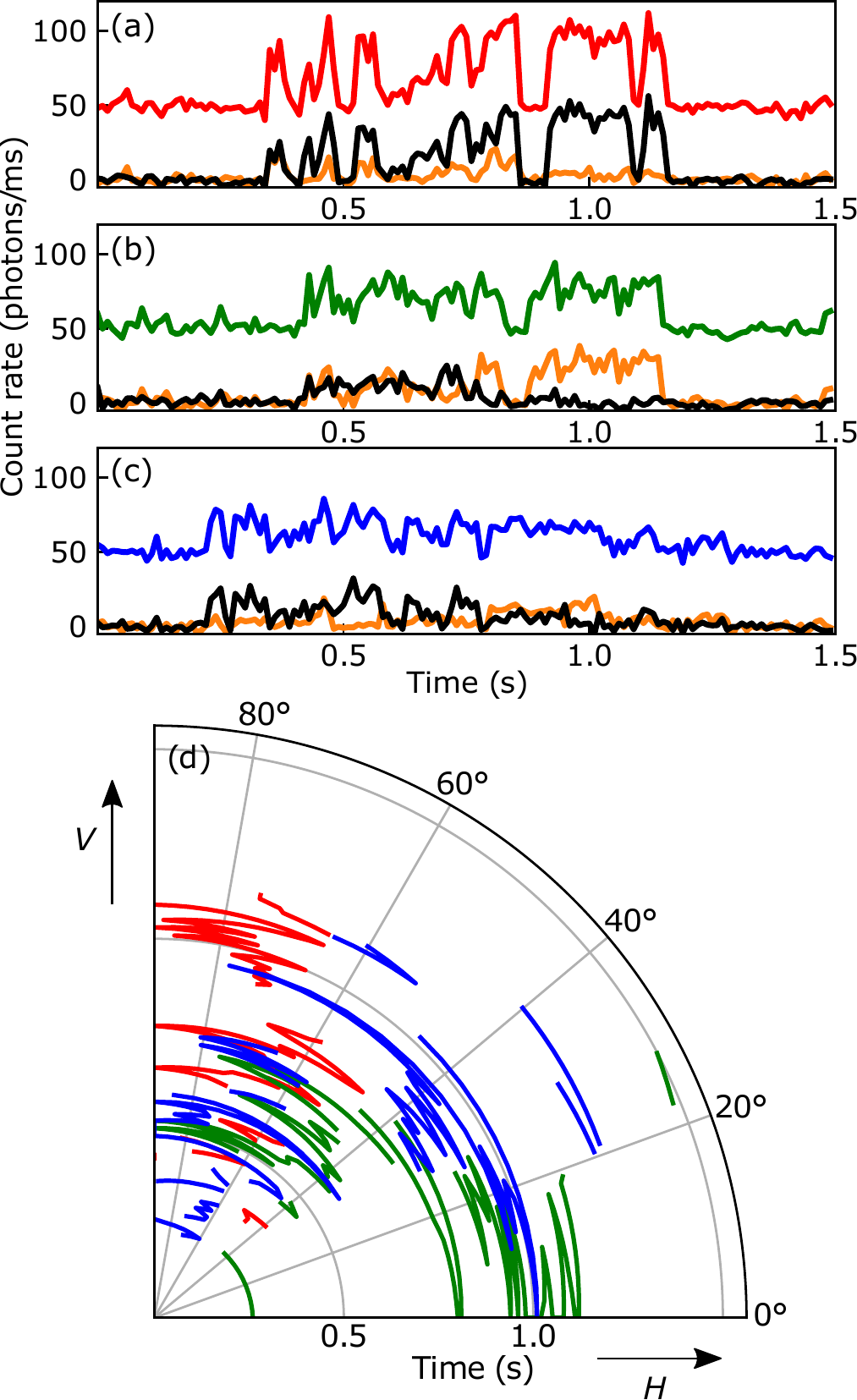}
\caption{(a)-(c) Fluorescence time traces of the horizontally (black) and vertically (orange) polarized components for three molecules. The red, blue and green lines represent the total intensities, added with a vertical offset. (d) Reconstructed dipole orientations of the three molecules as functions of time. Color codes are the same as in (a)-(c).}
\label{fig_hv}
\end{figure}

In addition to the translational diffusion, the rotational diffusion of single molecules on the surface can be tracked through polarization anisotropy measurements\,\cite{Lakowicz2007book}. We limit our studies to the two-dimensional (2D) rotational diffusion\,\cite{Harms1999} by deducing the 2D fluorescence polarization anisotropy $r=(I_h-I_v)/(I_h+I_v)$ for each molecule. Here, $I_{h,v}$ denote intensities of the horizontally and vertically polarized fluorescence light measured by the camera, respectively. The molecules are illuminated with circularly polarized light by adjusting the half- and quarter waveplates [see Fig.\,\ref{fig_setup}(c)]. To access these two polarization components, a Wollaston prism (WPM10, Thorlabs Inc.) is placed before the imaging lens [L2 in Fig.\ref{fig_setup}(c)], and separates the two polarization components by approximately $1\degree\,20^{'}$ in the horizontal direction. After the imaging lens, the two beams form two well-separated images on the sCMOS camera. To quantify the fluorescence anisotropy of individual molecules, we first sum up the two sub-images for each acquired frame after calibrating their relative pixel displacement. Molecules in each frame are identified from the summed images. The black (orange) lines in Fig.\,\ref{fig_hv}(a)-(c) display the time traces of $I_{h}$ ($I_{v}$) for three molecules. The total fluorescence intensities $I_{h}+I_{v}$ of the three molecules are represented by the red, green and blue lines, with a vertical offset added for clarity. Diffusion of their emission dipole orientation is evident through the change in the intensity ratio of the two polarization components. The random drops of intensities in the fluorescence time traces are due to photoblinking, {where the molecule enters a dark state such as the triplet or charge-separated state\cite{Zondervan2003,Zondervan2004}.} Photoblinking can last for milliseconds due to the lack of quenching agents in oxygen-free conditions\,\cite{Huebner2001, Renn2006}. Using the fluorescence signal, we reconstruct the transition dipole orientation $\theta$ of the molecules through the relation $\tan^2\theta=I_v/I_h$. The results are displayed in Fig\,\ref{fig_hv}(d). Here, we forced the angle to fall into the range of $[0, \pi/2]$. The discontinuities in the angular trajectories are results of photoblinking. Similar to the observations in translational diffusion, we observed molecules with strong angular confinement, as exemplified by the red curve. Molecules of this type are likely undergoing `wobbling-in-cone' type of rotation\,\cite{KinositaJr1982}. At the same time, we also observed molecules with higher rotational mobility, as indicated by the blue and green curves.

This set of measurements were performed with an exposure time of 10\,ms for each frame. Using this setting, an average uncertainty of $\pm8\degree$ in determining $\theta$ is achieved. The main contributions to the uncertainty come from photon shot noise and fluctuations of the fluorescence background. We also note that the polarization contrast of the setup is limited by transmission through the high-NA objective to the level of 15:1. The detection optical path features a slight dichroism, which amounts to about 3\% difference in transmission for the two orthogonal polarizations. This systematic offset is corrected in the data processing. 

{The inhomogeneity in the diffusional behaviors points to the co-existence of different adsorption states of perylene molecules on the surface. This may not be surprising considering the size of the molecule and the complexity of the amorphous fused silica surface. In early studies, it was identified that the freely vibrating hydroxyl groups are mainly responsible for the adsorption of smaller aromatic hydrocarbons such as benzene\cite{Hair1969}. However, other possible ending groups such as siloxane, and hydrogenbonded hydroxyl groups also affect the adsorption\cite{Hair1969}. In addition, surface defects could also introduce strong modifications of local potentials and lead to strong bindings.}

\section*{Temporal dynamics of fluorescence signal}
\begin{figure}[ht]
\centering\includegraphics[width=16cm]{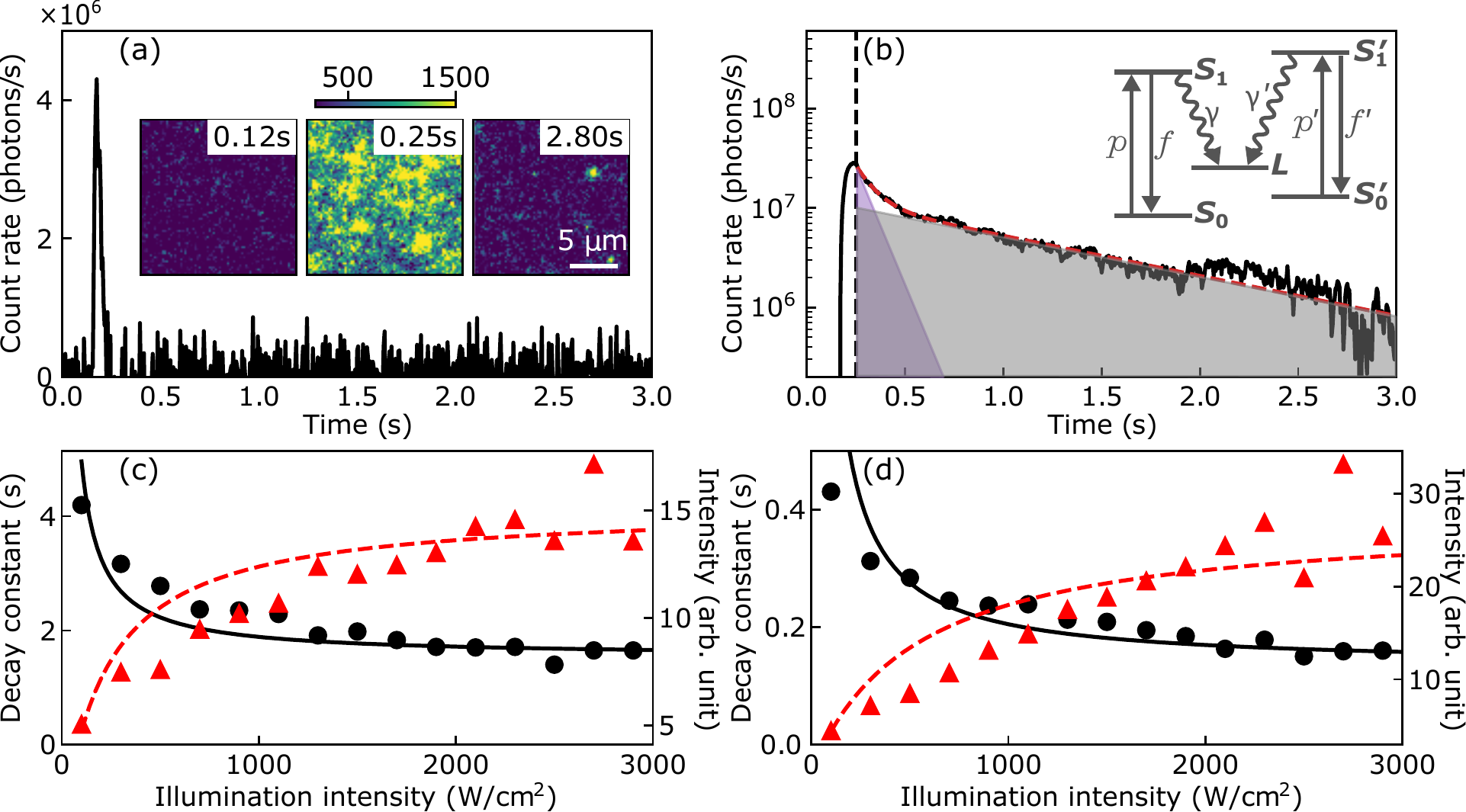}
\caption{(a) Change of fluorescence intensities between subsequent camera frames. The peak indicates the influx of molecules on the detection window. {The insets display fluorescence images of the observation area at different times. The colorbar has unit of photons per pixel and per second.} (b) Decay of the fluorescence intensity. The dashed red line represents a fit with a biexponential function. Contributions of the two components are displayed by the pink and grey shaded areas. The dashed black line indicates the end of molecule influx. {(c), (d) Time constants and amplitudes of the two exponential components as functions of illumination intensity. The black dots and curve indicate the measured and fitted time constants, respectively. The red triangles and dashed curve represent the same for the amplitudes.}}
\label{fig_adsorption}
\end{figure}

To gain further insight into this system, we measure the temporal dynamics of the ensemble fluorescence intensity under constant laser illumination. The measurement starts with the ablation of molecules from the target using a single laser pulse. After the ablation, the incidence of molecules on the fused silica window is measured through the change of fluorescence intensity between subsequent camera frames, as displayed in Fig.\,\ref{fig_adsorption}(a). The peak between 0.15\,s and 0.25\,s shows the time of arrival of molecules. {The insets display a sequence of images acquired at different times before and after the influx of molecules.} After the ablation, the window is continuously illuminated by the detection laser. Rapid decay of the fluorescence intensity is observed, as shown by the solid black curve in Fig.\,\ref{fig_adsorption}(b). {At longer time scales, the fluorescence intensity approaches the background level, as indicated by the last image in the inset of Fig.\,\ref{fig_adsorption}(a). A detailed observation of the decay signal reveals its non-mono-exponential character. Non-exponential decays can result from the heterogeneity in adsorption or internal states of molecules\cite{Zondervan2003}, or due to nonuniform illumination profiles\cite{Berglund2004}. The influence of the latter is minimized by micro-sectioning the camera images to areas much smaller than the illumination profile.} 

{The signal could be fit with a biexponential function, as shown by the dashed red line in Fig.\,\ref{fig_adsorption}(b). The amplitude and time constant of the two exponential components are plotted as functions of illumination intensities and shown in Figs.\,\ref{fig_adsorption}(c,d). Both time constants reduce with increasing illumination intensities, and clear saturation behaviors are observed. Two mechanisms can contribute to the decay of the fluorescence intensity, namely photobleaching and laser-induced desorption\cite{Arrowsmith1988,Meijer1990}. Polycyclic aromatic hydrocarbons such as perylene are considered highly photostable\cite{Basche2008_book}. Although their main photobleaching mechanism was attributed to photo-induced reactions in the presence of oxygen molecules\,\cite{Christ2001,Renn2006,Nothaft2012}, the exact pathways leading to photobleaching remain mostly unknown. A quantitative understanding of the laser-induced desorption would also require systematic {\it ab initio} calculations and complimentary experimental verifications. Despite these complexities, we attempt to understand the observations using a minimalistic model involving two adsorption states and five energy levels. As illustrated in the inset of Fig.\,\ref{fig_adsorption}(b), for each adsorption state, we consider the optical transitions between the singlet ground state $S_0$ ($S'_0$) to a singlet excited state $S_1$ ($S'_1$). Once a molecule is excited, it either returns to the ground state or reaches a dark state $L$ through photobleaching or desorption. We note that for both mechanisms there can be intermediate states involved. Here, for the sake of simplicity, we describe the contribution of all the pathways with an effective rate $\gamma$ ($\gamma'$). The rate of laser excitation and that of the decay via $S_1\rightarrow S_0$ ($S'_1\rightarrow S'_0$) are described by $p$ ($p'$) and $f$ ($f'$), respectively. The kinetics of the system can be described with rate equations based on this model. Knowing that the timescales of optical excitation and fluorescence decay are significantly shorter than that of photobleaching and desorption, we apply the steady-state approximation to the rate equations associated with $S_{0,1}$ ($S'_{0,1}$). This leads to the functional relation for the fluorescence intensity with time $I(t)=A{p}\exp{\left(-\frac{p}{p+f}\gamma t\right)}/({p+f})+A'{p'}\exp{\left(-\frac{p'}{p'+f'}\gamma' t\right)}/({p'+f'})$, with $A$ and $A'$ denoting the amplitudes of the two exponential components.}

{We fit the measured data in Figs.\,\ref{fig_adsorption}(c, d) with this relation. The results for the decay constants and their associated amplitudes are displayed as the black and red lines, respectively. For the time constants, although minor deviations of the fit curves are present, the general trends of saturation can be well recovered. The fit let us deduce $1/\gamma=1.56\,s$ and $1/\gamma'=0.14\,s$. Compared to the time constants, the amplitudes of the exponential components show larger deviations, especially at elevated laser intensities. The main reason is that the fluorescence intensities are dependent on the molecule density on the surface, which can vary in between measurements.}

\section*{Conclusion and outlook}
In conclusion, we demonstrated that by using an oil-immersion objective to image the inner surface of a home-made thin vacuum window, the adsorption, translation and rotational diffusion of single molecules on a surface under UHV can be studied with high spatial and temporal resolutions. {The dispersed diffusional dynamics suggest the co-existence of different adsorption states of perylene molecules. This is further supported by the temporal dynamics of the ensemble fluorescence intensity.} {Uncovering the full complexity of the system would require future work to combine {\it ab initio} calculations with complementary measurement techniques.} Along this line, the fused silica surface can be encapsulated with thin layers suited for surface science studies. {For example, transparent dielectric layers could be incorporated through standard deposition or substrate transfer techniques. It should also be possible to apply conductive layers without significantly impeding the optical transmission. For example, conductive oxides such as indium tin oxide have a transmission of about 80\% in the spectral range of 400\,nm-600\,nm even with 500\,nm layer thickness\,\cite{Prepelita2017}. Gold layer with a thickness of 2\,nm still permits optical transmission of $60\%$ in the same wavelength range\cite{Axelevitch2012}. Two-dimensional materials, e.g., graphene introduces only about 2\% absorption per monolayer.} These will make the platform compatible with scanning probe techniques, which could enable direct imaging of the adsorption sites with atomic scale resolutions\cite{Schuler2013,Wong2007,Civita2020}. Such a system would also facilitate simultaneous correlative electron and light microscopy\cite{Liv2013,Zonnevylle2013}. Alternatively, implementing gas phase photoionization detection schemes will permit independent determinations of the residence times and scattering velocities of desorbed molecules\cite{Harding2017}. Lastly, the bonded window can be attached to a UHV flange through glass-to-metal bonding, to replace the rubber seal. This should permit higher bakeout temperatures and lower vacuum levels down to the $10^{-11}$ mbar range.

\begin{acknowledgement}

The authors thank Kilian Singer, Zhipeng Huang and Benjamin Stickler for fruitful discussions and feedback to the manuscript. This work is supported by Deutsche Forschungsgemeinschaft (DFG, German Research Foundation) through CRC 1319.

\end{acknowledgement}



\makeatletter
\providecommand{\doi}
  {\begingroup\let\do\@makeother\dospecials
  \catcode`\{=1 \catcode`\}=2 \doi@aux}
\providecommand{\doi@aux}[1]{\endgroup\texttt{#1}}
\makeatother
\providecommand*\mcitethebibliography{\thebibliography}
\csname @ifundefined\endcsname{endmcitethebibliography}
  {\let\endmcitethebibliography\endthebibliography}{}

\end{document}